\documentclass[%
 reprint,
 amsmath,amssymb,
 pra,
]{revtex4-2}
\usepackage{hyperref}
\usepackage{graphicx}
\usepackage{dcolumn}
\usepackage{bm}
\usepackage{color,soul}
\usepackage{amsmath}
\usepackage{braket}
\usepackage{subcaption}
\usepackage{siunitx}
\DeclareSIUnit{\calorie}{cal}
\DeclareSIUnit{\Calorie}{\kilo\calorie}
\DeclareSIUnit{\au}{{a.u.}}
\DeclareSIUnit{\aut}{{AUT}}

\begin{document}

\preprint{APS/123-QED}

\title{An Open Quantum Systems approach to proton tunnelling in DNA}

\author{Louie Slocombe}
\email{l.slocombe@surrey.ac.uk}
\affiliation{
	Leverhulme Quantum Biology Doctoral Training Centre,
	University of Surrey,
	Guildford, GU2 7XH, UK.}
\author{Marco Sacchi}
\email{m.sacchi@surrey.ac.uk}
\affiliation{
	Department of Chemistry,
	University of Surrey,
	Guildford, GU2 7XH, UK.}
\author{Jim Al-Khalili}
\email{j.al-khalili@surrey.ac.uk}
\affiliation{
	Department of Physics,
	University of Surrey,
	Guildford, GU2 7XH, UK.
}

\date{\today}

\begin{abstract}
One of the most important topics in molecular biology is the genetic stability of DNA. One threat to this stability is proton transfer along the hydrogen bonds of DNA that could lead to tautomerisation, hence creating point mutations. We present a theoretical analysis of the hydrogen bonds between the Guanine-Cytosine (G-C) nucleotide, which includes an accurate model of the structure of the base pairs, the quantum dynamics of the hydrogen bond proton, and the influence of the decoherent and dissipative cellular environment. We determine that the quantum tunnelling contribution to the process is several orders of magnitude larger than the contribution from classical over-the-barrier hopping. Due to this significant quantum contribution, we find that the canonical and tautomeric forms of G-C inter-convert over timescales far shorter than biological ones and hence thermal equilibrium is rapidly reached. Furthermore, we find a large tautomeric occupation probability of $1.73\times 10^{-4}$, suggesting that such proton transfer may well play a far more important role in DNA mutation than has hitherto been suggested. Our results could have far-reaching consequences for current models of genetic mutations.
\end{abstract}


\maketitle

In their seminal paper, Watson and Crick proposed that the tautomerisation of DNA base pairs could produce stable errors in the genetic code \cite{Watson01011953}. The proposed mechanism for such tautomerisation is double proton transfer along the hydrogen bonds within base pairs, G-C or A-T \cite{RN181, RN603}. This mechanism is of particular interest due to what has been predicted to be a significant contribution from the quantum tunnelling of the protons through a potential barrier separating the nucleotide base pairs on the two strands of DNA \cite{Slocombe2021Quantum}. The mechanism has been widely studied theoretically \cite{brovarets2019atomistic,godbeer2015modelling,Slocombe2021Quantum,gheorghiu2020influence} and, more rarely, experimentally \cite{pohl2018proton,kimsey2018dynamic}.

\begin{figure}
\begin{subfigure}{\columnwidth}
  \includegraphics[width=\columnwidth]{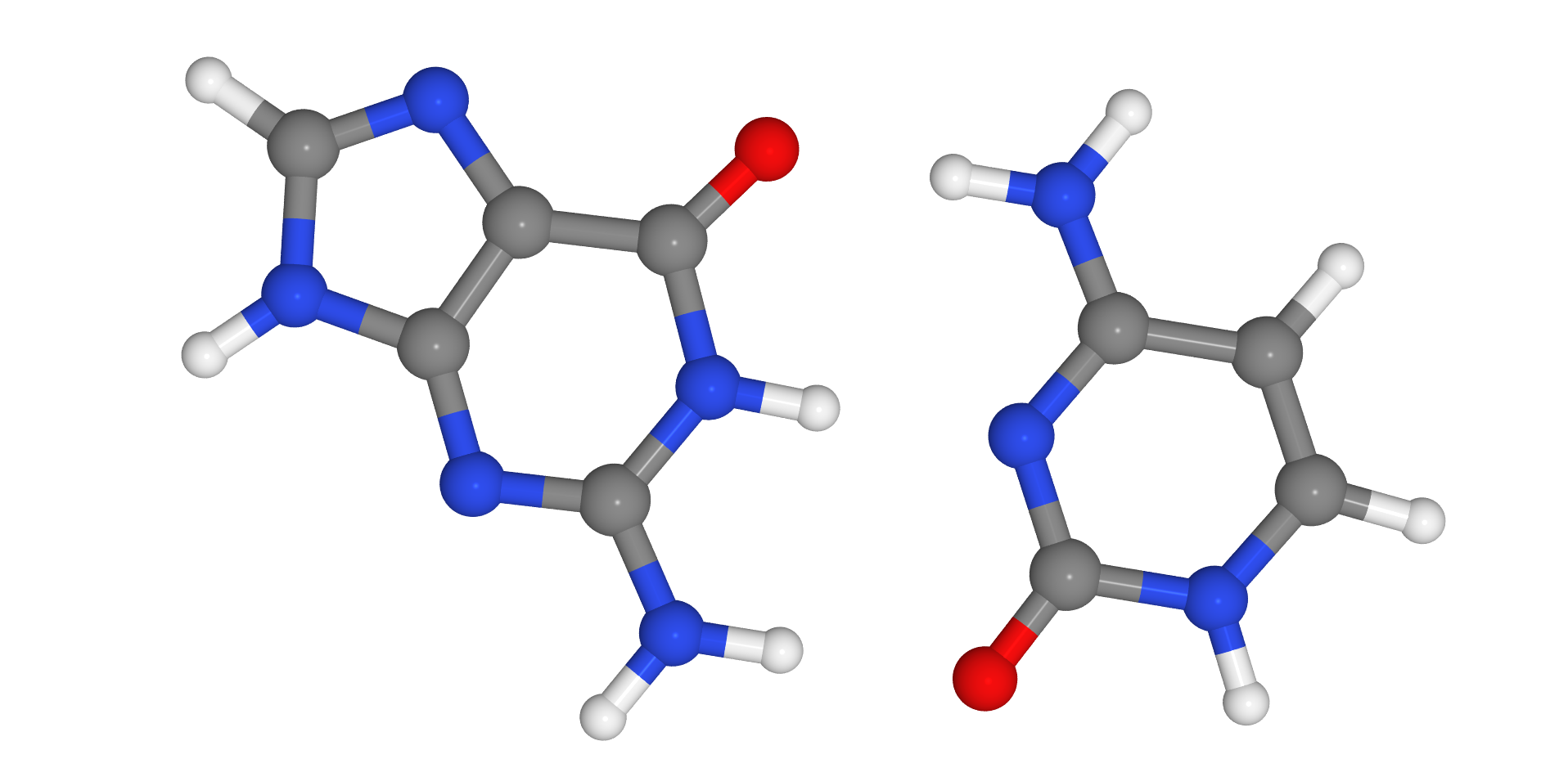}
  \caption{Canonical standard form of G-C}
\end{subfigure}

\begin{subfigure}{\columnwidth}
  \includegraphics[width=\columnwidth]{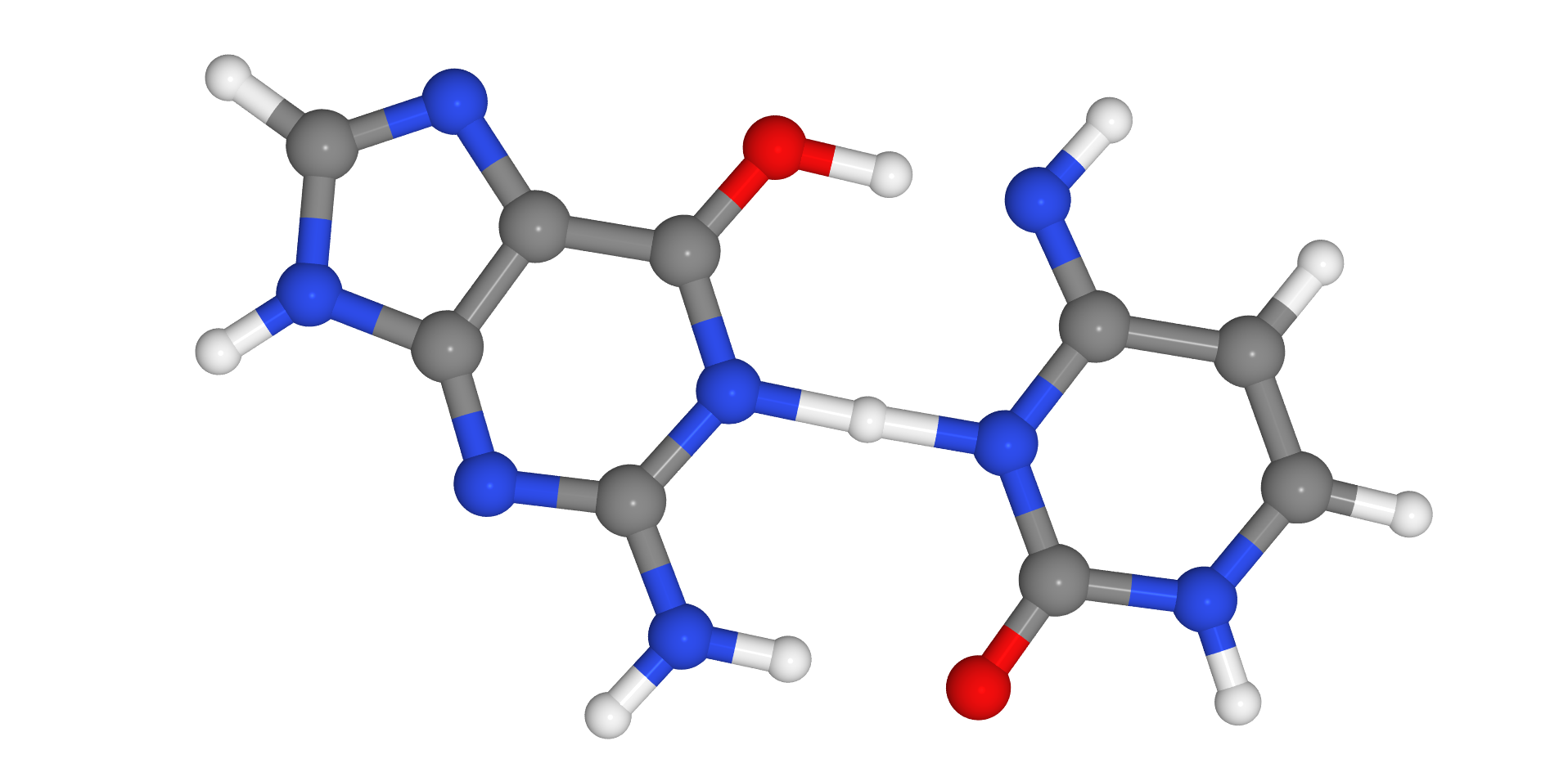}
  \caption{Energetic maximum along the reaction coordinate.} 
\end{subfigure}

\begin{subfigure}{\columnwidth}
  \includegraphics[width=\columnwidth]{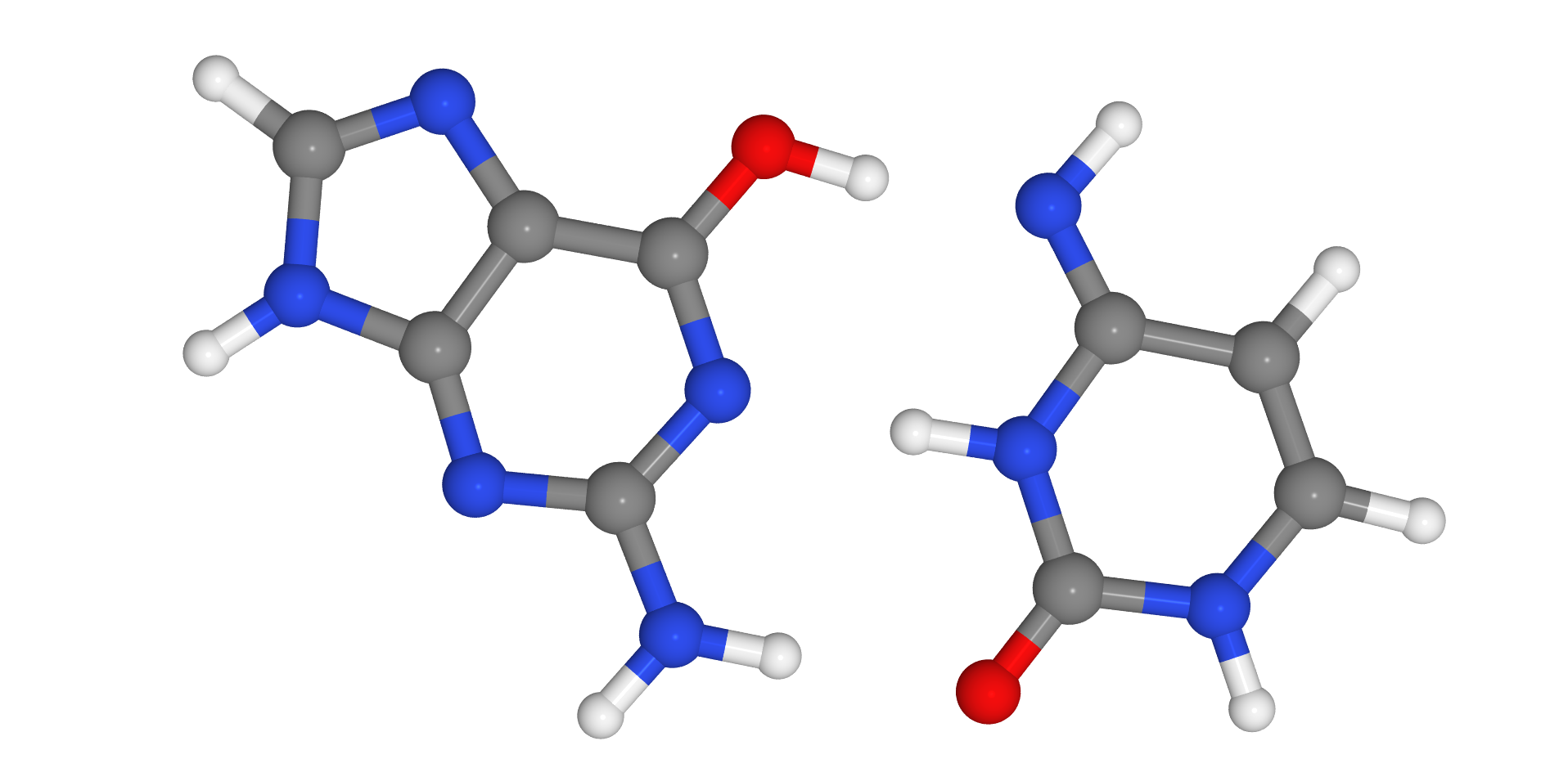}
  \caption{Tautomeric non-standard G*-C*}
\end{subfigure}

\caption{Schematic depiction of the G-C proton transfer reaction. The starred forms labels correspond to the tautomeric (enol and imino) forms of the DNA base.\label{fig:lowdin}}
\end{figure}

When the H-bond protons transfer across from a base site on one strand to the corresponding site on the other strand, each base changes from its standard canonical to its tautomeric form (see Fig.~\ref{fig:lowdin}). If this tautomeric pair can survive the DNA cleavage process in the helicase (an enzyme that aids DNA strand separation), then each strand could pass through the DNA replication machinery (the replisome), whereby the tautomeric form of the base is mismatched with the wrong corresponding base on the copy strand \cite{Watson01011953,kim2021quantum}. Furthermore, this mismatched pair can evade fidelity check-points of the replisome by adopting a structure similar to a Watson and Crick base pair \cite{kim2021quantum,brovarets2019atomistic}, resulting in an erroneous mismatch and hence a point mutation (for example, G-C $\Leftrightarrow$ G*-C* $\Rightarrow$ G*-T). 

The role of the double proton transfer mechanism has been hotly debated for over half a century, with several authors \cite{brovarets2019atomistic,brovarets2014tautomerization,brovarets2014can,gheorghiu2020influence,soler2019proton} suggesting that it does not meet the criteria needed to survive the DNA replication process, sine the tautomers would need to survive the nanosecond timescale of the Helicase cleavage process \cite{brovarets2019atomistic,brovarets2014tautomerization,brovarets2014can}. We present here a different and, we believe, more convincing argument. We show that the double proton transfer process takes place so quickly (in comparison with biologically relevant timescales) that the lifetimes of the tautomeric states do not play a part in determining the likelihood of a base pair mismatch. Rather, it is the ratio between the concentration of tautomeric base pairs to canonical base pairs present at chemical equilibrium that matters.

We assume, as previously observed \cite{godbeer2015modelling,perez2010enol}, that the two protons undergo asynchronous transfer, suggesting that the rate-determining process is the transfer of a single proton across the barrier, with the other proton then moving in the opposite direction to preserve electroneutrality. Thus, we need only consider the single proton transfer process between the G-C base pair in this work. The thermodynamic equilibrium constant, $K_{\textrm{eq}}$, is generally written as $K_{\textrm{eq}} =$ [G*-C*]/[G-C] (the ratio between the concentration of products -- tautomeric pair -- and reactants -- canonical pair -- present at chemical equilibrium). However, this definition assumes perfect thermodynamic equilibrium, which is not necessarily true. More correctly, an activity coefficient, $\Gamma$, needs to be included as a factor such that $K_{\textrm{eq}} = \Gamma\times$[G*-C*]/[G-C]. For example, quantum mechanical effects such as thermally enhanced tunnelling and zero-point energy can lead to an activity coefficient that is far from unity \cite{cramer2013essentials}.

For G-C tautomerism,  $K_{\textrm{eq}}$ can be readily obtained, either from the difference in Gibbs free energy, $\Delta G$, between products and reactants via $K_{\textrm{eq}}= \exp \left( -\Delta G / k_\textrm{b} T \right)$ and calculated with \textit{ab initio} methods, or measured experimentally. However, without an accurate estimate of the activity coefficient, we cannot determine the quantity of interest, [G*-C*]/[G-C]. Instead, we carry out a fully quantum mechanical calculation to obtain the G*-C* occupation probability within an open quantum system approach. We find it to be several orders of magnitude greater than the observed rate ($10^{-8}$ per base pair \cite{wu2017dna}) of spontaneous mutations through, for example, copying errors, suggesting that tautomerisation may well play an important role in point mutations.

Much recent theoretical effort has been devoted to investigating tautomerisation of A-T and G-C base pairs via double proton transfer, such as the use of path integral molecular dynamics \cite{fang2016inverse,pohl2018proton,perez2010enol}. Applying these methods suggests that protons delocalisation causes the strength of the hydrogen bonds in DNA base pairs to follow an anomalous temperature dependence due to the subtle balance of several quantum effects involved in the binding. The problem with such approaches is that while the proton dynamics are treated quantum mechanically, the coupling between them and the surrounding environment is dealt with via approximate ``classical" thermostats. However, it is well known that crucial quantum effects such as decoherence and dissipation are not treated correctly in adiabatic, thermally uncoupled models of tunnelling \cite{liu2021understanding,godbeer2015modelling}.
Therefore, we employ in this work a fully open quantum system treatment of the proton transfer mechanism to model the G-C tautomerisation. Our decision to study G-C rather than A-T is because the tautomeric state, A*-T*, is extremely unstable, with a fleetingly short lifetime, and therefore not as likely to play a role in mutagenesis \cite{Slocombe2021Quantum,gheorghiu2020influence,brovarets2019atomistic,soler2019proton}. 

We divide the problem into two. First, we obtain an accurate description of the potential energy landscape of the double H-bond between the C and G bases; next we model the proton dynamics using an open quantum systems (OQS) model to account for dissipation and decoherence effects due to the surrounding cellular environment.

\begin{figure}
\centering
\vspace*{0.5cm}
\includegraphics[width=\columnwidth]{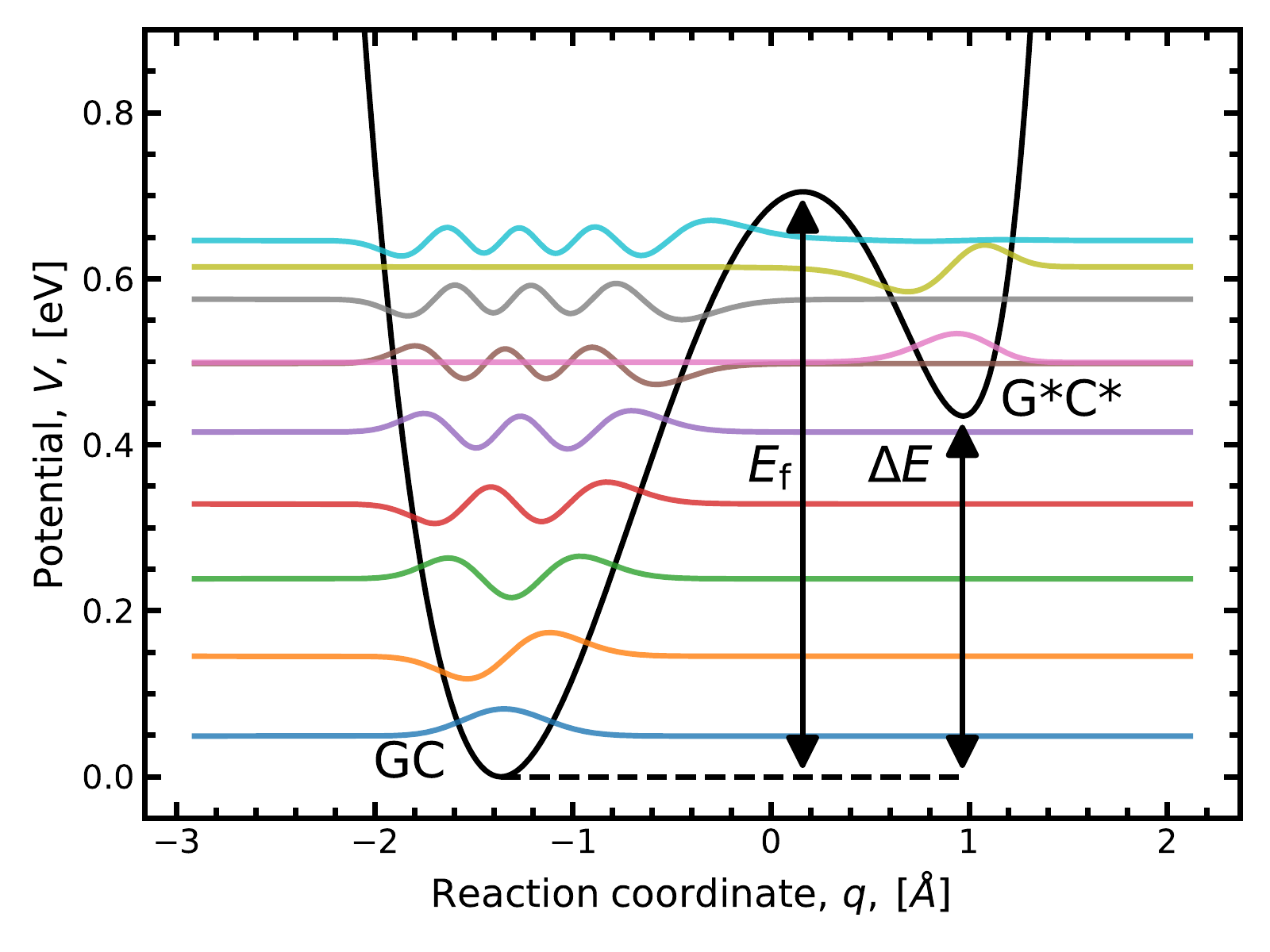}
\caption{G-C double proton transfer energy landscape converted into a back-to-back double Morse potential. The first ten eigenstates are displayed. With values $E_{0} = \SI{0.049}{\eV}, \; E_\mathrm{f}=\SI{0.705}{\eV}, \; \Delta E=\SI{0.435}{\eV}$. \label{fig:gc_pot}}
\end{figure}

The proton transfer process is initially described by projecting the reaction pathway on a generalised reaction coordinate that connects the canonical to the tautomeric state via a single transition state \cite{wales2003energy}\cite{Slocombe2021Quantum} on a chemically accurate potential energy surface (PES) in the form of an asymmetric double-well potential (Fig.~\ref{fig:gc_pot}, see Method section). The parameters of this potential are used to calculate the classical rate contribution within Eyring's framework, which is outlined in the Methods section. Thus, the reaction is characterised by a $\Delta E$ (energetic asymmetry between the bottoms of the two potential wells) of \SI{0.435}{\eV}, a high forward barrier of \SI{0.704}{\eV} and a reverse barrier of \SI{0.270}{\eV}. The PES allows us to calculate the quantum tunnelling correction and estimate the lifetime of the tautomeric state. Fig.~\ref{fig:gc_pot} also shows the first ten energy eigenvalues for a single proton in this potential, six of which are entirely local to the G-C canonical configuration, with the seventh eigenstate the first to have a non-zero amplitude in the shallow (tautomeric) well. 

\section*{The system-bath model} 
The potential is used in the Hamiltonian of our OQS master equation in which the dynamics of the proton transfer are described following the Caldeira and Leggett quantum Brownian motion model \cite{caldeira1983path}, in which the quantum system (a proton in the double-well potential) interacts with a surrounding infinite thermal bath of harmonic oscillators (representing the cellular environment). The bath degrees of freedom are then integrated over using the path integral formalism of Feynman and Vernon \cite{feynman2000theory}. 

The equivalent phase-space formulation of the Caldeira and Leggett  master equation, also known as Wigner–Moyal-Caldeira-Leggett (WM-CL) equation, is written as \cite{Wigner1932Quantum,caldeira1983path}
\begin{widetext}
\begin{equation} \label{eq:WM-CL}
\frac{\partial W}{\partial t}=
 \underbrace{
-\frac{p}{m} \frac{\partial W}{\partial q}
+\frac{\partial V}{\partial q} \frac{\partial W}{\partial p}
-\frac{\hbar^{2}}{24} \frac{\partial^{3} V}{\partial q^{3}} \frac{\partial^{3} W}{\partial p^{3}} +\mathcal O\left(\hbar^{4}\right)
}_{\text{Schr{\"o}dinger dynamics}}
 \underbrace{
+\gamma \frac{\partial p W}{\partial p}
}_{\text{Dissipation}}
\underbrace{
+\gamma m k_{B} T \frac{\partial^{2} W}{\partial p^{2}}
}_{\text{Decoherence}},
\end{equation}
\end{widetext}
and describes the time evolution of the Wigner function, $W(q,p,t)$. The initial state is taken to be a non-equilibrium distribution localised in the reactant well.
\begin{equation}
\begin{split}
W(q, p, t=0)=
&\frac{1}{\mathcal{N}} 
\exp \left(-\mathcal{H}/ (2 E_{\omega}) \right)\\
& (1-\hat{h}(q)) \exp \left(-\mathcal{H} / (2 E_{\omega}) \right),
\end{split}
\end{equation}
where $\mathcal{H} = p^2 / (2m) + V(q)$, and $\mathcal{N}$ is a normalisation constant. 

In Eq.~(\ref{eq:WM-CL}), the phenomenological friction constant, $\gamma$, describes the strength of the coupling to the surrounding (ohmic) heat bath \cite{caldeira1983path}, $k_{\mathrm{B}}$ is Boltzmann's constant, $T$ is the bath temperature, and the quantum `system of interest' (the H-bond proton) has position $q$, momentum $p$ and mass $m$. Eq.~(\ref{eq:WM-CL}) also contains two terms describing the dissipation and decoherence that arise from the coupling of the proton with the bath. It is a deterministic dynamical equation that is rigorously valid only in the weak-coupling regime and in the high-temperature limit \cite{caldeira1983path}. 

\begin{figure}
\centering
\includegraphics[width=\columnwidth]{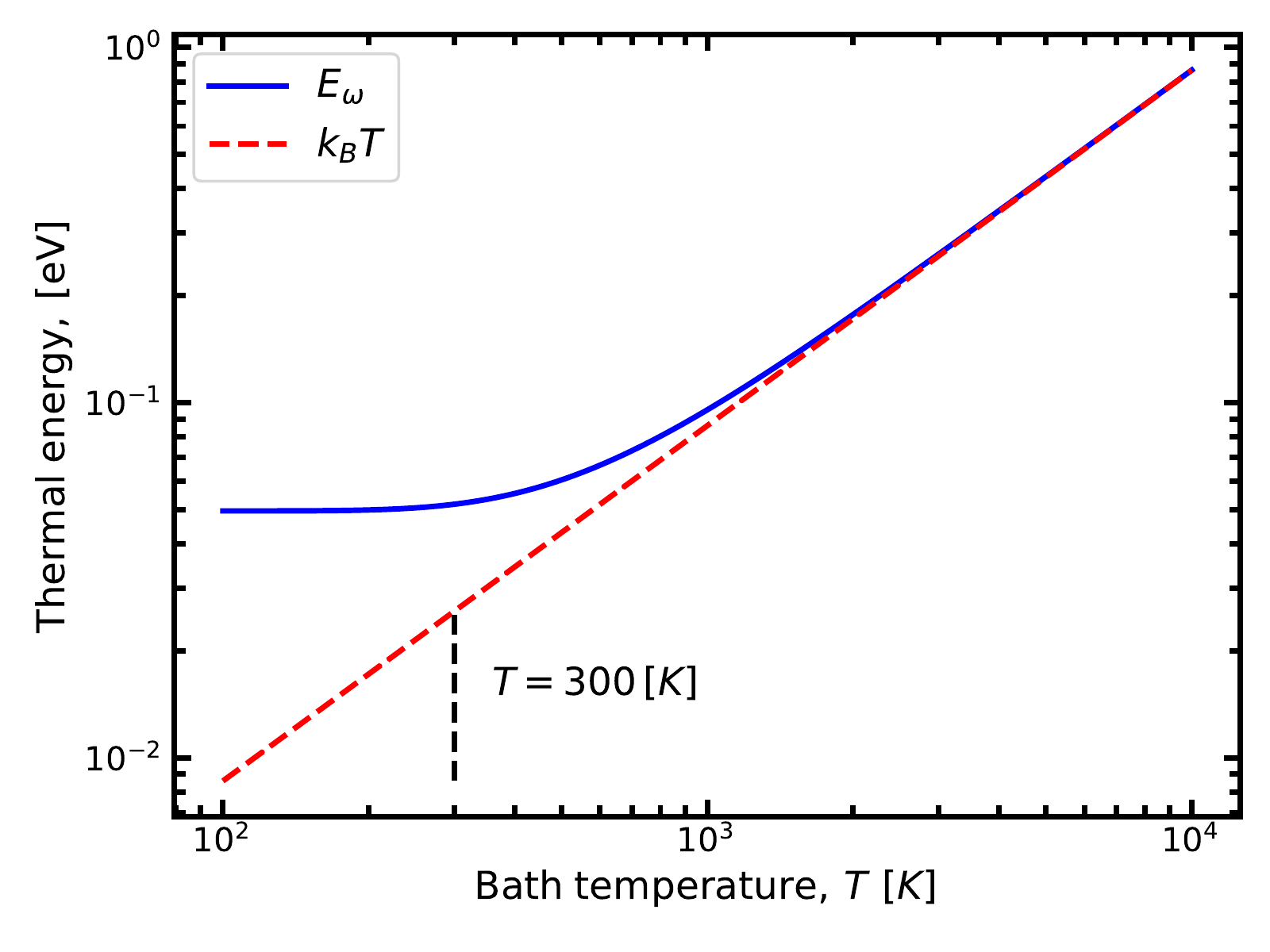}
\caption{The breakdown of the high-temperature approximation, taking the leading order term of $\operatorname{coth}$, (the dashed red line) compared to the low-temperature correction (the unbroken blue line). \label{fig:gc_low_temp_limit}}
\end{figure}

However, for biologically relevant bath temperatures ($T\approx \SI{300}{\kelvin}$), the high-temperature limit is not a valid approximation. Consequently, our model requires a low-temperature correction applied in Eq.~(\ref{eq:WM-CL}) whereby the thermal energy is replaced with the zero-point energy \cite{burghardt2002quantum,agarwal1971brownian,haake1985strong,haake1986master,hughes2005dissipative},  
\begin{equation} \label{eq:thermal_correction}
k_{\mathrm{B}} T \rightarrow E_{\omega}=\frac{\hbar \omega}{2} \operatorname{coth}\left(\frac{\hbar \omega}{2 k_{B} T}\right),
\end{equation}
where the high temperature limit is simply the leading term in the Taylor expansion of $\operatorname{coth} x=\frac{1}{x}+\frac{x}{3}-\frac{x^{3}}{45}+\cdots$ \cite{caldeira1983path,burghardt2002quantum,agarwal1971brownian,haake1985strong,haake1986master,hughes2005dissipative}. Here, $\omega$ is the frequency of the lowest energy eigenstate in the double-well potential. Making the replacement described in Eq.~(\ref{eq:thermal_correction}) permits the use of the WM-CL equation at lower temperatures, in which $\omega$ can be approximated by the second derivative of the potential around its global minimum. Fig.~\ref{fig:gc_low_temp_limit} shows the breakdown of the high-temperature approximation for the G-C proton transfer potential. Further discussion on the low-temperature correction and its justification can be found in the Methods section. To account for this, we solve Eq.~(\ref{eq:WM-CL}) with the factor $k_{\mathrm{B}}T$ in the final term replaced by the right hand side of Eq.~(\ref{eq:thermal_correction}).

Assuming that the bath is dominated by the thermal fluctuations of the surrounding water molecules, we can estimate the value of $\gamma$. Water has a collision spectrum in the range $3300 - 3900 \, \SI{}{\per \cm}$ \cite{gottwald2015applicability}. 

\section*{Lifetime of the Tautomer}
In order to determine whether or not tautomerisation plays an important role in biology, we need to know the lifetime of the tautomeric state. Therefore, we calculate the quantum contribution, $k_{\mathrm{QM}}$, to the chemical reaction rate by monitoring how the flux of the probability changes between the left and right-hand well \cite{tanimura1991quantum,tanimura1992interplay,zhang2020proton,ishizaki2005multidimensional,montgomery1979trajectory}. For a short time, we expect transient behaviour of the density with re-crossings of the transition state \cite{hanggi1990reaction} rather than exponential macroscopic decay of flux passing through the barrier. The phenomenological rate law applies only after a characteristic time $\tau_c$ in which this transient phase has ended. 

The tunnelling factor, $\kappa$, calculated in the Methods section, is found to be very large ($\sim 10^5$) indicating that the quantum contribution to the reaction rate is non-trivial. With this value, we obtain a forward and reverse reaction rate of $k_{\mathrm{f}}=\SI{7.61e5}{\per\second}$ and  $k_{\mathrm{r}}=\SI{1.69e13}{\per\second}$ respectively. Whereas the half-life of the reactant and product is now $\SI{9.11e-7}{\second}$ and $\SI{4.09e-14}{\second}$ respectively. 

Comparatively, setting the quantum contribution in Eq.~\ref{eq:trans_rate} to unity gives a classical forward and reverse reaction rate of
$k_{\mathrm{f}}=\SI{7.596}{\per\second}$ and $k_{\mathrm{r}}=\SI{1.692e8}{\per\second}$. The classical equilibrium constant, $K_{\mathrm{eq}}$, is on the order of $10^{-8}$ which is in agreement with Gheorghiu \textit{et al.} \cite{gheorghiu2020influence}. Whereas the half-life of the reactant and product is $\SI{0.0913}{\second}$ and $\SI{4.1e-9}{\second}$ respectively.

These quantum corrected half-lives are significantly shorter than the classical estimates, indicating the protons are sufficiently delocalised along the hydrogen bond and populate both states; the forward and reverse reaction rates are so quick that the system continuously converts between the canonical and tautomeric forms. This fast interconversion timescale suggests that an equilibrium description can be safely adopted.

\section*{Tautomeric state occupation probability}
This timescale of the proton transfer ($\SI{10}{\femto\second}-\SI{100}{\nano\second}$) is significantly shorter than the cleavage timescale in the helicase ($t \gg \SI{100}{\nano\second}$). Therefore, the G-C base pair enters the active site of the helicase in thermal equilibrium between canonical and tautomeric states with its eigenstates populated according to a Maxwell-Boltzmann distribution, $W^{\mathrm{eq}} (q, p) = W(q, p, t \rightarrow \infty)$. 

\begin{figure}
\centering
  \includegraphics[width=\columnwidth]{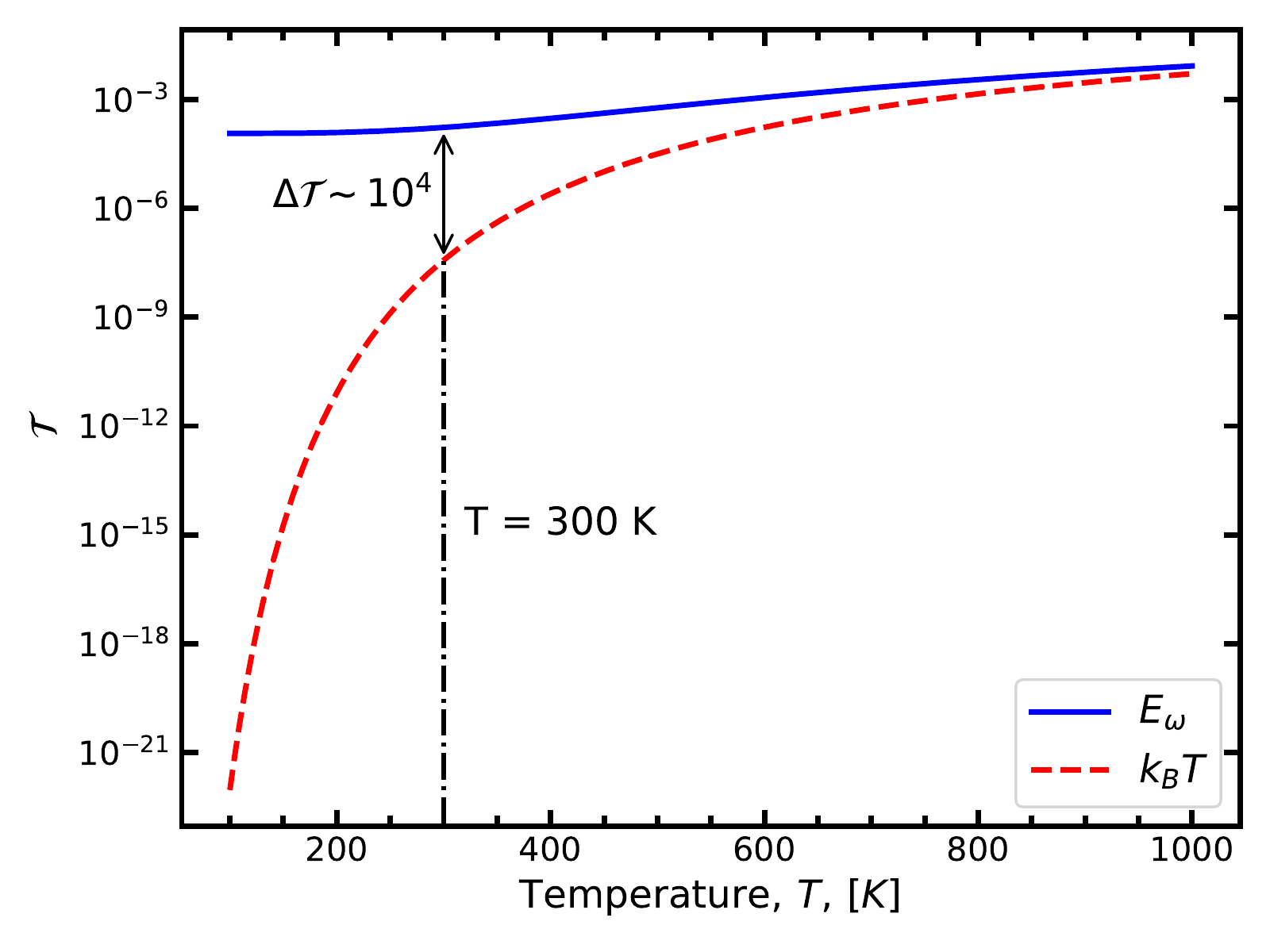}

\caption{The tautomer occupation probability is plotted as a function of bath temperature. The low-temperature corrected result is shown as the solid line in blue and the high-temperature approximation as a dashed red line. \label{fig:ic_temp_dep}}
\end{figure}

The tautomer occupation probability can now be calculated by integrating the Wigner function at equilibrium over all momenta, $p$, and over position, $q$, spanning only the width of the shallow well (the product side of the barrier),
\begin{equation} \label{eq:tun_prob}
\mathcal{T}=\int\!\!\! \int \hat{h}(q) W^{\mathrm{eq}}(q, p) \; dq \, dp\ ,
\end{equation}
where $\hat{h}(q)$ is a Heaviside step function that projects onto the product side of a transition state dividing surface (reaction barrier). The top panel of Fig.~\ref{fig:ic_temp_dep} examines the probability of finding the thermal distribution in the tautomeric well. At $\SI{300}{\kelvin}$ we observe an tautomer occupation probability of $1.73 \times 10^{-4}$. At biological temperature, the probability in the right hand well for the $E_{\omega}$ curve levels out as the zero-point energy of the oscillator dominates. On the other hand, if we neglect the low-temperature correction, the probability rapidly becomes negligible as the temperature drops. Incorporating the low-temperature correction shows that tunnelling can play a role even with little bath excitation due to zero-point energy corrections. 

As is clear from the annotations on Fig.~\ref{fig:ic_temp_dep}, there is a factor of $10^4$ difference between the thermalised tautomer occupation probability obtained from the two temperature treatments. 

While the surrounding environment quickly causes dissipation and decoherence of the quantum system (the transferring proton), it also initially provides a vital source of thermal activation, exciting the proton to higher energy eigenstates in the double-well that can promote tunnelling through to the tautomeric side and leaving a non-trivial population in this state once it reaches thermal equilibrium.

\section*{Conclusion}
We have explored the proton transfer mechanism in the H-bonds of the G-C base pair within an open quantum system model.  In this approach, the quantum system (the H-bond proton in the double-well potential) undergoes dissipation and decoherence due to coupling to a surrounding heat bath (the cellular environment). The environment also acts as a source of thermal activation, exciting the proton to higher energy states that promote tunnelling across to the right-hand tautomeric state. At $t\rightarrow\infty$, the thermal equilibrium distribution gives a probability of $1.73 \times 10^{-4}$ for the proton being in the tautomeric form, which is more than four orders of magnitude larger than that predicted by classical and semiclassical studies previously reported for this system.

Purely classical calculations predict that the tautomeric state is rarely populated due to its high forward reaction barrier. Furthermore, the relatively small reverse reaction barrier suggests that the tautomer is classically unstable with a lifetime of the order Helicase cleavage timescale. However, in our open quantum system approach, the tunnelling factor obtained is on the order of $10^{5}$, suggesting that the system readily interconverts between the canonical and tautomeric configuration via quantum effects. 

Previously, it was suggested by Brovarets' \textit{et al.} \cite{brovarets2019atomistic} and Gheorghiu \textit{et al.} \cite{gheorghiu2020influence} that if the tautomeric lifetime is much shorter than helicase cleavage timescale, little to no product would survive the process. Our investigations determined that the quantum rate is significantly higher than the classical rate for a wide range of bath coupling strengths. The tautomer's lifetime calculated with quantum correction suggests that equilibrium concentration of this species is
quickly reached. Furthermore, find that the forward and reverse proton transfer processes are significantly quicker than the expected helicase cleavage timescale. Because of this effective equilibrium in the helicase active site, we can adopt the tautomer occupation probability as a metric to define the ratio of canonical to tautomeric. 

Attempting to identifying potential kinetic isotope effect in this system is problematic since investigating biological assays in deuterated solvents comes with a whole host of problems, for example, viscosity effects inhibiting correct protein function \cite{hohlefelder2013heavy}. In addition, the impairment of gene expression at the transcription or translation level could preclude the measurement of the KIE discussed in the Methods section. Therefore, more direct measurements via nuclear magnetic resonance shifts, such as those applied to the G-T wobble mismatch, need to be applied \cite{kimsey2018dynamic}. 

On the other hand, computational studies suggest that the monomeric form of the tautomers are stable and extremely long-lived due to their prohibitively large reaction barriers \cite{Slocombe2021Quantum,soler2019proton,gorb1999theoretical,podolyan2003ab}. Based on previous evaluations \cite{Slocombe2021Quantum} of tunnelling in this system, we expect that the tautomeric population of the monomeric forms are unlikely to see any meaningful change while in transit from the helicase until it matches another base pair. Consequently, any last line of defence against the tautomer's uptake would take place during the exonuclease proofreading process during which slight modifications to the DNA structure caused by the tautomeric mismatch could be corrected.

Experimental tautomerisation rates for G-C base pairs are still not available, but we nevertheless believe that our theoretical results should radically revise our understanding of the likelihood of point mutations in DNA, and we hope that they will encourage new experimental measurements. One can note that our model predicts a rate of tautomerisation much higher than the overall rate of spontaneous mutations ($\sim 10^{-8}$) \cite{wu2017dna}, but this is consistent with the well-known presence of highly efficient DNA repair mechanisms. It is also difficult to comment on the overall efficiency of these repair mechanisms because other spontaneous mutations will also take place in DNA aside from tautomerisation \cite{kunkel1986base,zahurancik2014significant}.

\section*{Methods}
\section*{Numerical Methods}
We solve Eq.~\ref{eq:WM-CL} numerically using the method of lines approach in which $q$ and $p$ are discretised to a fixed equally spaced lattice with $N_q$ points in range $[q_{\mathrm{min}}, \, q_{\mathrm{max}}]$ and $N_p$ equally spaced points in range $[p_{\mathrm{min}}, \, p_{\mathrm{max}}]$. The partial derivatives in space are expanded using a second-order central finite difference approach. The outer coefficients are set to zero, corresponding to Dirichlet (reflecting) boundary conditions. To integrate the equations in time, we utilise the VCABM5 algorithm \cite{butcher2000numerical}, an adaptive 5th order Adams-Moulton method implemented in the DifferentialEquations.jl ecosystem \cite{rackauckas2017differentialequations}.  We find that VCABM5 offers a good trade-off between accuracy and speed. To simplify the numerical calculations, we drop the third-order potential terms in Eq.~\ref{eq:WM-CL}. We set the mass to a proton $m=\SI{1836}{\au}$, in contact with a thermal bath with $T=\SI{298.15}{\kelvin}$.

\section*{The model potential}
Table~\ref{tab:gc_params} contains all the parameters defining the potential corresponding to the tautomerisation reaction of G-C. 

\begin{table}[h]
\caption{The potential parameters in Hartree atomic units defining the G-C double proton transfer reaction.}\label{tab:gc_params}
\begin{ruledtabular}
\begin{tabular}{lcc}
Parameter & Symbol & Value \\ 
\colrule
Canonical well energy & $V_{1}$ & $\SI{0.1617}{\hartree}$ \\
Tautomeric well energy & $V_{2}$ & $\SI{0.082}{\hartree}$ \\

Canonical width parameter & $a_{1}$ & $\SI{0.305}{\bohr}$ \\
Tautomeric width parameter & $a_{2}$ & $\SI{0.755}{\bohr}$ \\

Canonical distance & $r_{1}$ & $\SI{-2.7}{\bohr}$ \\
Tautomeric distance & $r_{1}$ & $\SI{2.1}{\bohr}$ \\

Forward reaction barrier & $E_{f}$ & $\SI{0.0259}{\hartree}$ \\
Reverse reaction barrier & $E_{r}$ & $\SI{0.00992}{\hartree}$ \\

Spring constant of the barrier & $\omega_b$ & $\SI{0.00277}{\au}$ \\
\end{tabular}
\end{ruledtabular}
\end{table}

\section*{The low-temperature limit}
\begin{figure}
\centering
\begin{subfigure}{\columnwidth}
  \centering
  \includegraphics[width=\columnwidth]{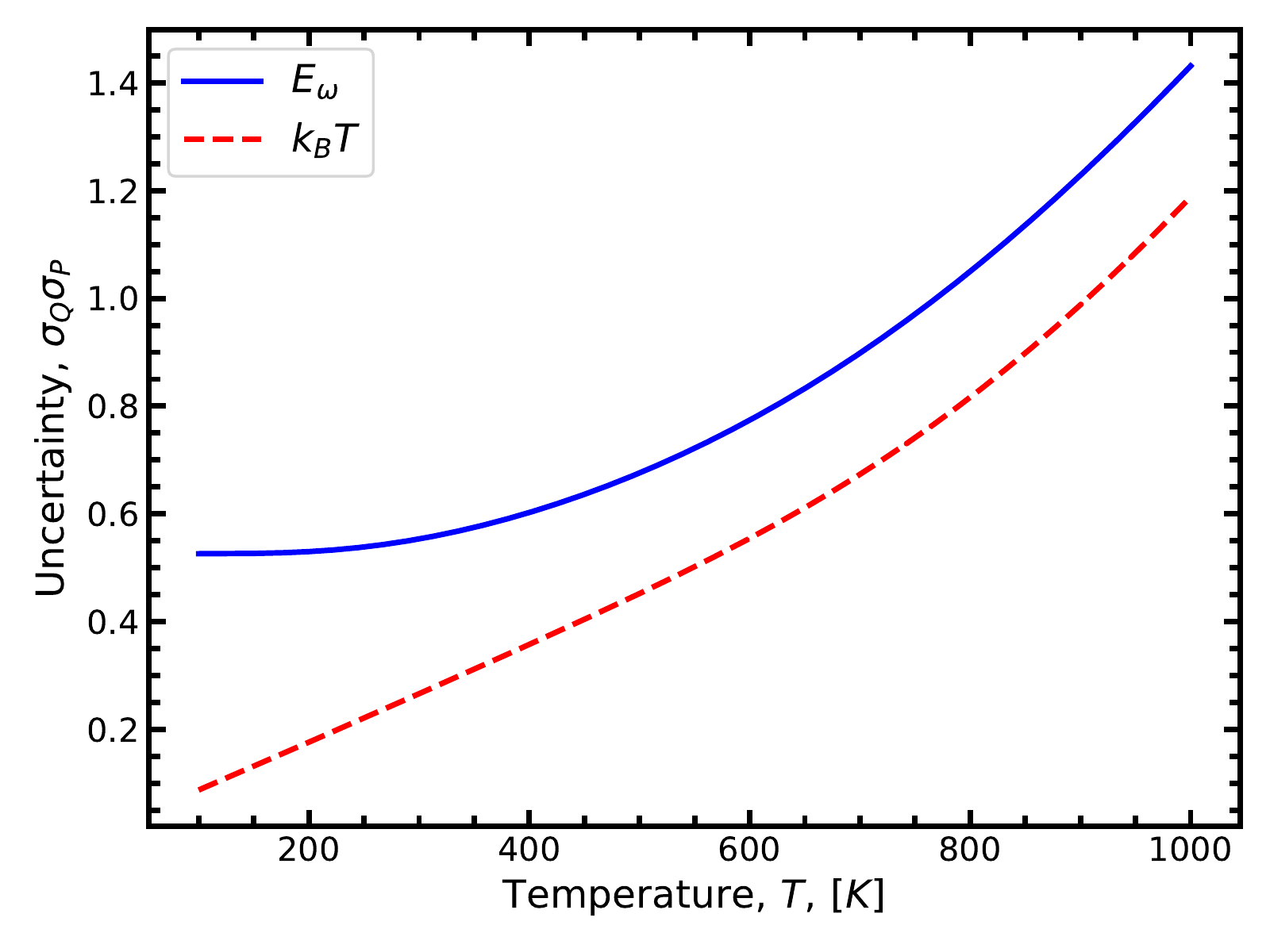}
  \caption{Position-momentum uncertainty}
\end{subfigure}

\begin{subfigure}{\columnwidth}
  \centering
  \includegraphics[width=\columnwidth]{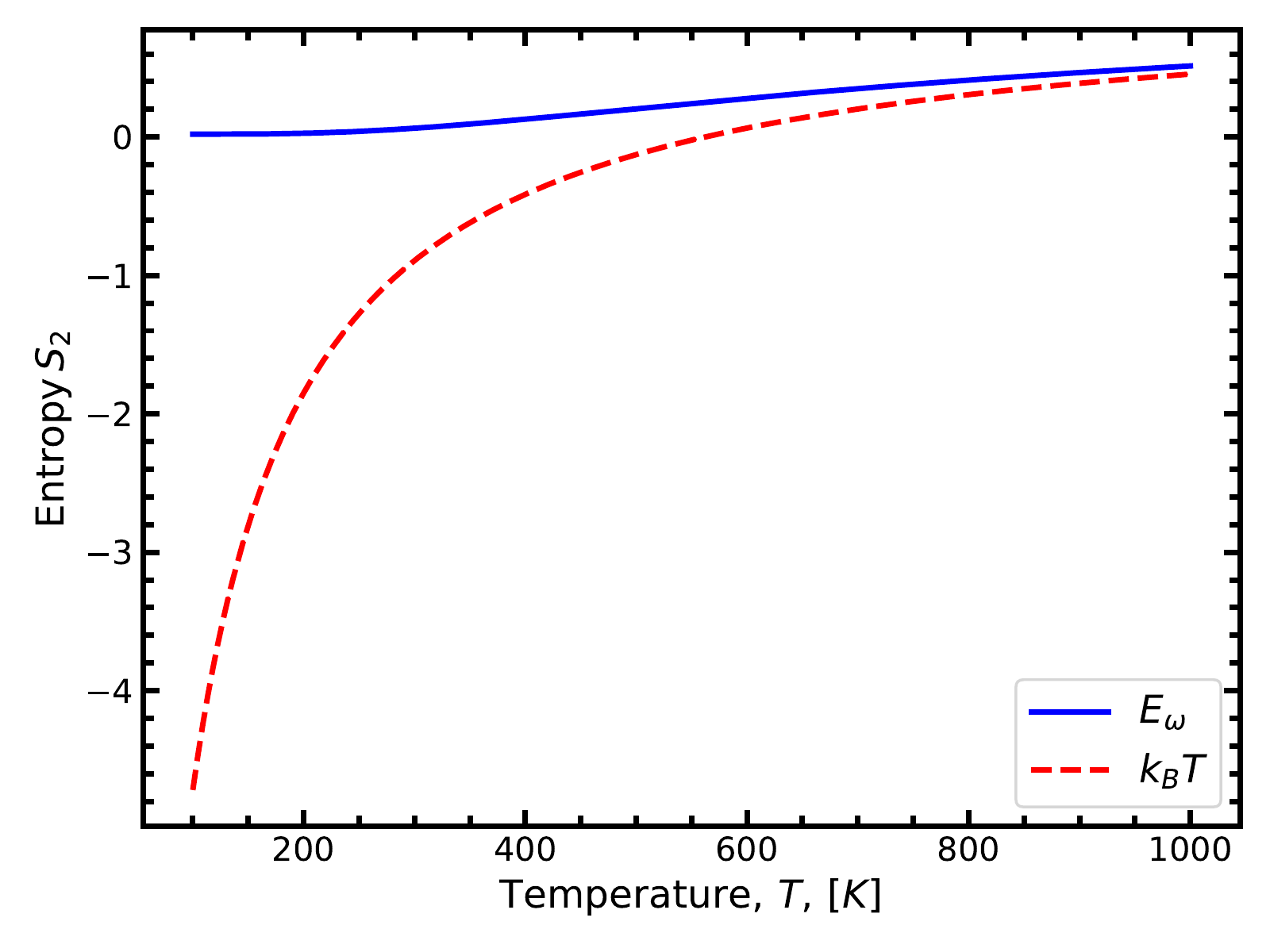}
  \caption{$S_{2}$ linear entropy}
\end{subfigure}

\caption{The thermal properties of the double-well potential are displayed as a function of the temperature correction. The low-temperature correction is shown as the solid line in blue, whereas the high-temperature approximation is shown in a dashed red line. \label{fig:ic_temp_dep_more}}
\end{figure}

To verify the validity of the low-temperature correction, we monitor some properties of the distribution (See Fig.~\ref{fig:ic_temp_dep_more}). Here the thermal properties can be obtained by integrating Eq.~\ref{eq:WM-CL} to $t\rightarrow \infty$ at a given temperature. In the high-temperature case, Eq.~\ref{eq:WM-CL} tends to a thermal average over the classical Hamiltonian. However, at low temperature, it is dominated by the quantum correction to the zero-point energy of the effective oscillator (Eq.~\ref{eq:thermal_correction}).

We note additionally that Heisenberg's position-momentum uncertainty principle must also be upheld,
\begin{equation} \label{eq:uncert_principle}
\sigma_{Q} \sigma_{P} \geq \frac{\hbar}{2}\ ,
\end{equation}
where 
\begin{equation}
\sigma_{Q}=\sqrt{\left\langle Q^{2} \right\rangle - \langle Q \rangle^{2}}, \; \;
\sigma_{P}=\sqrt{\left\langle P^{2} \right\rangle - \langle P \rangle^{2}}.
\end{equation}
Within the atomic unit system, a minimally uncertain distribution will take on a value of $0.5$, with a value less than this violating the uncertainty principle. The middle panel of Fig.~\ref{fig:ic_temp_dep_more} demonstrates that the high-temperature treatment becomes unphysical below $\sim \SI{600}{\kelvin}$, whereby the distribution becomes increasingly point particle-like, violating the uncertainty principle. At low temperature, the linear trend continues to the origin, implying that at $\SI{0}{\kelvin}$, the distribution is perfectly point-like, with zero uncertainty. On the other hand, incorporating the correction prevents the compression of the distribution in phase space at lower temperatures - demonstrating an asymptotic convergence to $0.5$. At much higher temperatures shown in the plot, the two approaches converge.  

If the system starts in a pure quantum state, the bath's interactions will decohere it by driving it to a mixed state \cite{zurek1993preferred}. The canonical quantum entropy via von Neumann, $S=-\operatorname{Tr}(\hat{\rho} \ln \hat{\rho})$ (where $\rho$ is the density matrix) offers a measure of the statistical uncertainty represented by a quantum state. The von Neumann entropy is generally regarded as the preferred choice of a metric of coherence. However, in phase-space negative values can occur in the logarithm, which presents a problem with its evaluation. As a consequence, the von Neumann is too restrictive for our needs. Alternatively, the linearised entropy is a computationally convenient but accurate approach to measure coherence \cite{zurek1993coherent,wlodarz2003entropy}. We can monitor the amount of decoherence using 
\begin{equation} \label{eq:entropy}
S_{2}= 1 - (2 \pi \hbar)^{D} \int \int W^{2}(q, p, t) \; dq \, dp.
\end{equation}
Here, $D$ corresponds to the number of degrees of freedom (unity in this case) \cite{zurek1993coherent,wlodarz2003entropy}. For a pure, unmixed state, the $S_{2}$ entropy takes on a value of zero. While the system evolves in contact with the bath, it begins to decohere, observed as an increase in linear entropy. If the bath coupling is turned off, we strictly observe no entropy change. As seen in the bottom panel of Fig.~\ref{fig:ic_temp_dep_more}, the breakdown of the high-temperature approach is demonstrated as negative entropy values below $\sim \SI{600}{\kelvin}$. Negative entropy implies that the system can be in a more ordered state than an unmixed, pure state, which is problematic. Whereas the low-temperature correction correctly implies $S_{2} \rightarrow 0$.

\section*{Chemical reaction rate}
In order to investigate the kinetics of the reaction, we need to determine the quantum contribution to the chemical reaction rate by monitoring the flux of the density passing through the transition state (barrier). 

The tunnelling factor, $\kappa$, can be calculated using the classical and quantum contribution to the rate,
\begin{equation}\label{eq:trans_rate}
\kappa(T)= 1 +
\underbrace{\beta \exp \left(\beta E_{\mathrm{f}} \right) 
}_{\mathrm{Classical}} \,
\underbrace{
\lim _{t \rightarrow \tau_c}
\frac{(d / dt) \delta N(t)}{\delta N(t)}
}_{\mathrm{Quantum}}.
\end{equation}

We obtain a quantum contribution, $k_{\mathrm{QM}}$, to the chemical reaction rate by monitoring the flux of the probability changes between the left and right-hand well \cite{tanimura1991quantum,tanimura1992interplay,zhang2020proton,ishizaki2005multidimensional,montgomery1979trajectory}. The change can be determined by,
\begin{equation}
\begin{split}
\delta N(t)= 
&\int \int
W \left(q, p; t \right)
\, \hat{h}(q) \, dp \, dq \\ 
&- \int  \int
W \left(q, p; t=0 \right)
\, \hat{h}(q) \, dp \, dq.
\end{split}
\end{equation}
Here, $\hat{h}(q)$ is the same Heaviside step function defined before, which delimits the left and right-hand well. The quantum contribution to the reaction rate is determined using Eq.~\ref{eq:trans_rate}
Hence, the full forward and reverse rate constants, $k_{\mathrm{f}}$ and $k_{\mathrm{r}}$, are obtained from,
\begin{equation}\label{eq:rate}
k_{\mathrm{f,r}} = \kappa(T) \frac{1}{h \beta} \exp \left(-\beta E_{\mathrm{f,r}}\right),
\end{equation}
A reaction timescale can be obtained by inspecting the inverse of Eq.\ref{eq:rate}.

\section*{Kinetic Isotope Effect}
We have also investigated the impact of the kinetic isotope effect as a function of temperature (doubling the proton's mass if replaced by a deuteron). A strong dependence of the reaction rate on the reduced mass of the system can indicate tunnelling. Hence the kinetic isotope can be used to probe if quantum effects dominate. Using the same parameters before but doubling the mass, we investigate the kinetic isotope effect (KIE).
\begin{equation}
    \mathrm{KIE} = \frac{k^{\mathrm{p}}_{\mathrm{QM}}}{k^{\mathrm{d}}_{\mathrm{QM}}},
\end{equation}
where $k^{\mathrm{p}}_{\mathrm{QM}}$ ($k^{\mathrm{d}}_{\mathrm{QM}}$) is the rate for a proton (deuteron). We neglect any changes to the transfer energy landscape due to isotopic substitutions, such as zero-point energy and free energy contributions, and consider only the effect on the reaction rate due to modifications of the mass terms in Eq.~\ref{eq:WM-CL}.

At low temperatures, the KIE rapidly increases, providing further evidence that the transfer process becomes increasingly dependent on the quantum (tunnelling) contribution. Meanwhile, at high temperatures, the KIE exponentially tails off. At biological temperatures the $\mathrm{KIE}= 30$.

\section*{Acknowledgements}
We are grateful for financial support from the Leverhulme Trust. In addition, we acknowledge helpful discussions with the members of the Leverhulme Quantum Biology Doctoral Training Centre. Particular thanks go to Johnjoe McFadden and Max Winokan, who both offered many productive conversations. Finally, the authors thank the University of Surrey for access to the Eureka HPC. This work used the \href{https://www.archer2.ac.uk}{ARCHER2} UK National Supercomputing Service via the UKCP consortium.

\section*{Data availability}
The data presented in the figures of this article are available from the corresponding authors upon reasonable request.

\section*{Code availability}
Numerical simulations were performed with Julia code that makes use of the differential equations package available on \href{https://github.com/SciML/DifferentialEquations.jl}{Github}. The Julia source codes are available from the corresponding authors upon reasonable request.

\section*{Author contributions}
M.S. and J.A-K. conceived and designed this research, L.S. built the computational apparatus. All the authors contributed to the preparation of the manuscript.

\section*{Competing interests}
The authors declare no competing interests.

\bibliography{references.bib}


\clearpage

\end{document}